# A sixth-order finite difference scheme with the minimized dispersion and adaptive dissipation for solving compressible flow

Youtian Su, Yanhui Li, Yu-Xin Ren*
School of Aerospace Engineering, Tsinghua University, Beijing 100084, China
*Corresponding author: ryx@tsinghua.edu.cn

**ABSTRACT**
  The dispersion and dissipation properties of a scheme are important to realize high-fidelity simulations of the compressible flow, especially the cases with broadband length scales. It has been recognized that the minimization of dispersion error is an effective method to improve the precision. In addition, the proper dissipation of a scheme is important to restrain the non-physics oscillations and reserve details of flows simultaneously. The authors have previously proposed a scale sensor to adjust the numerical dissipation of a fourth-order finite difference scheme according to the local scale of the flow. In this paper, the scale sensor is further modified for the sixth-order finite difference scheme to achieve minimized dispersion and adaptive dissipation properties. Firstly, the scale sensor quantifies the local length scale of the numerical solution as the effective scaled wavenumber. Then, the dispersion-dissipation condition is used to construct the relationship between the dissipation/dispersion parameter and the effective scaled wavenumber. Therefore, a sixth-order finite difference scheme with minimized dispersion and adaptive dissipation (MDAD6th) is proposed. Several benchmark test cases with broadband length scales are presented to clarify the high resolution of the new scheme.

*Keywords:* Low dispersion scheme; Adaptive dissipation scheme; Scale sensor; Approximate dispersion relation

## 1. Construction of MDAD6th

For the spatial derivative at the grid point j, analogous to the finite volume method, numerical approximation is achieved with the utilization of the flux at $j-\frac{1}{2}$ and $j+\frac{1}{2}$.

$$\left.\frac{\partial f}{\partial x}\right|_{x_j} \approx f_\Delta(x_j) = \frac{\hat{f}_{j+\frac{1}{2}} - \hat{f}_{j-\frac{1}{2}}}{\Delta x} \quad (1)$$

Now considering the numerical flux $\hat{f}_{j+\frac{1}{2}}$ at $j+\frac{1}{2}$, to realize the degree of freedom for the adaption, MDAD6th (Sun et al. 2014)(Li, Chen, and Ren n.d.) divides 8 points into five groups of candidate stencils (figure 1).

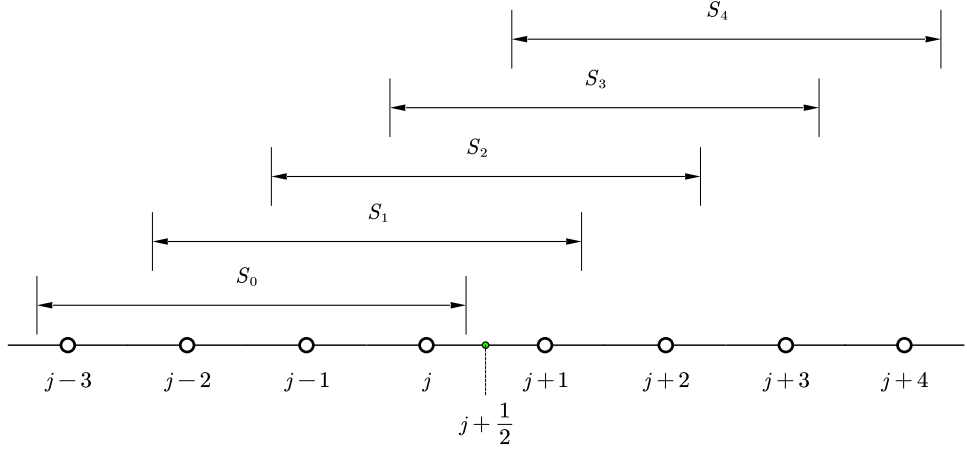

**Figure 1:** five stencils of MDAD6th with the use of 8 candidate points.

Specifically, for each candidate stencil $S_i$, we can derive the highest-precision scheme $q_i$ of the numerical flux that can be achieved at $j+\frac{1}{2}$ (eq. 2).

$$\begin{cases} q_0 = -\frac{1}{4}f_{j-3} + \frac{13}{12}f_{j-2} - \frac{23}{12}f_{j-1} + \frac{25}{12}f_j \\ q_1 = \frac{1}{12}f_{j-2} - \frac{5}{12}f_{j-1} + \frac{13}{12}f_j + \frac{1}{4}f_{j+1} \\ q_2 = -\frac{1}{12}f_{j-1} + \frac{7}{12}f_j + \frac{7}{12}f_{j+1} - \frac{1}{12}f_{j+2} \\ q_3 = \frac{1}{4}f_j + \frac{13}{12}f_{j+1} - \frac{5}{12}f_{j+2} + \frac{1}{12}f_{j+3} \\ q_4 = \frac{25}{12}f_{j+1} - \frac{23}{12}f_{j+2} + \frac{13}{12}f_{j+3} - \frac{1}{4}f_{j+4} \end{cases} \quad (2)$$

The complete version of the MDAD scheme is a linear combination of the numerical flux of each stencil. The construction process realizes the adaptation of dispersion and dissipation at the expense of two-order precision in exchange for two independent parameters $\gamma_{disp}$ and $\gamma_{diss}$ that control the dispersion and dissipation. In conclusion, we use eight points to obtain a six-order scheme with two-degree-of-freedom.

*Table 1*

| $C_0$ | $C_1$ | $C_2$ | $C_3$ | $C_4$ |
|---|---|---|---|---|
| $2\gamma_{diss} - 2\gamma_{disp}$ | $\frac{1}{5} + 16\gamma_{diss} - 4\gamma_{disp}$ | $\frac{3}{5} + 12\gamma_{disp}$ | $\frac{1}{5} - 16\gamma_{diss} - 4\gamma_{disp}$ | $-2\gamma_{diss} - 2\gamma_{disp}$ |

Eq. (3) gives the expression of the numerical flux at $j+\frac{1}{2}$ and the parameters are from table (1),

$$\hat{f}_{j+\frac{1}{2}} = \sum_{i} C_i \cdot q_i \tag{3}$$

## 2. Analysis of Dissipative Properties of MDAD

As for a one-dimensional wave $Exp(ikx)$, the analytical solution should be $i \cdot k \cdot Exp(ikx)$. However, as shown in Eq. (4), due to the truncation error produced by the specific finite difference scheme of Eq. (1), the calculation result will be different.

$$f_\Delta(x_j) = i \cdot k' \cdot Exp(ikx) = i \cdot \left( \frac{\Re(k \cdot \Delta x) + i \cdot \Im(k \cdot \Delta x)}{\Delta x} \right) \cdot Exp(ikx) \tag{4}$$

The functions $\Re(k)$ and $\Im(k)$ related to the equivalent wavenumber $k'$ correspond to the dispersion and dissipation properties of the semi-discrete scheme respectively. Eq. (5) gives the expression of MDAD's functions $\Re(k)$ and $\Im(k)$, ($k \cdot \Delta x = \varphi_0$, representing the phase change across each grid),

$$\begin{cases} \Re(\varphi_0) = \frac{3}{2}\sin(\varphi_0) - 14\gamma_{disp}\sin(\varphi_0) - \frac{3}{10}\sin(2\varphi_0) + 14\gamma_{disp}\sin(2\varphi_0) \\ \qquad\qquad + \frac{1}{30}\sin(3\varphi_0) - 6\gamma_{disp}\sin(3\varphi_0) + \gamma_{disp}\sin(4\varphi_0) \\ \Im(\varphi_0) = \gamma_{diss} \cdot [-35 + 56\cos(\varphi_0) - 28\cos(2\varphi_0) + 8\cos(3\varphi_0) - \cos(4\varphi_0)] \end{cases} \tag{5}$$

It can be seen that the dispersion properties and the dissipation properties are related to the parameters $\gamma_{disp}$ and $\gamma_{diss}$ independently.

## 3. Construction of Scale Sensors

### 3.1. Scale Sensors based on the ratio of derivatives

We assumed that the scale corresponds to the spatial frequency of a characteristic sine wave of the flow field information contained in the eight candidate points. The corresponding scale in the flow field of a standard trigonometric function ($f = \sin(kx + \phi)$) should itself be the spatial frequency of the sine wave. Using the derivation property of the trigonometric function, for the absolute value of the different-order derivative $f^{(i)}_{j+\frac{1}{2}}$ at the point $x_{j+\frac{1}{2}}$ calculated by the finite difference scheme $f_i$ with the grid points, the ratio should meet the properties in Eq. (6),

$$\begin{cases} \frac{\partial f}{\partial x}\bigg|_{x_{j+\frac{1}{2}}} = k \cdot \cos(kx+\phi) \\ \frac{\partial^2 f}{\partial x^2}\bigg|_{x_{j+\frac{1}{2}}} = -k^2 \cdot \sin(kx+\phi) \\ \frac{\partial^3 f}{\partial x^3}\bigg|_{x_{j+\frac{1}{2}}} = -k^3 \cdot \cos(kx+\phi) \\ \frac{\partial^4 f}{\partial x^4}\bigg|_{x_{j+\frac{1}{2}}} = k^4 \cdot \sin(kx+\phi) \\ \frac{\partial^5 f}{\partial x^5}\bigg|_{x_{j+\frac{1}{2}}} = k^5 \cdot \cos(kx+\phi) \end{cases} \quad \begin{cases} f^{(1)}_{j+\frac{1}{2}} = \left|\frac{f_1}{\Delta x}\right| \\ f^{(2)}_{j+\frac{1}{2}} = \left|\frac{f_2}{\Delta x^2}\right| \\ f^{(3)}_{j+\frac{1}{2}} = \left|\frac{f_3}{\Delta x^3}\right| \\ f^{(4)}_{j+\frac{1}{2}} = \left|\frac{f_4}{\Delta x^4}\right| \\ f^{(5)}_{j+\frac{1}{2}} = \left|\frac{f_5}{\Delta x^5}\right| \end{cases} \quad (6)$$

In the further construction, if the truncation error is not considered, the ratio of the derivative function of different orders satisfy Eq. (7),

$$k = \sqrt{\frac{f^{(i+2)}}{f^{(i)}}} \approx \sqrt{\frac{f^{(i+2)}_{j+\frac{1}{2}}}{f^{(i)}_{j+\frac{1}{2}}}} = \sqrt{\frac{f_{i+2}}{f_i}} \cdot \frac{1}{\Delta x}$$

$$\varphi_i \equiv \sqrt{\frac{f_{i+2}}{f_i}} \approx \varphi_0 \quad (7)$$

Due to the existence of rounding errors in the actual calculation, the relative error will increase sharply when the formula takes the extreme value of the corresponding order derivative of the trigonometric function. In order to ensure that the scale sensors can work normally in the special phase, the method of adding the ratio of different-order derivatives to the denominator is adopted in this paper. Specifically, there are two construction methods (Eq. 8).

$$\begin{aligned} sensor 82 core: \quad & \varphi_{pre} = \sqrt{\frac{f_3+f_4}{f_1+f_2}} \\ sensor 83 core: \quad & \varphi_{pre} = \sqrt{\frac{f_3+f_4+f_5}{f_1+f_2+f_3}} \end{aligned} \quad (8)$$

However, due to the truncation error, in fact, for the highest difference accuracy $\varphi_i$ (i = 1,2,3), the corresponding prediction scale results are different (figure 2). Furthermore, the mathematical meaning of the scale calculated by the scale sensor and the legality of the post-processing theory is not clear. The details are shown in the figure below,

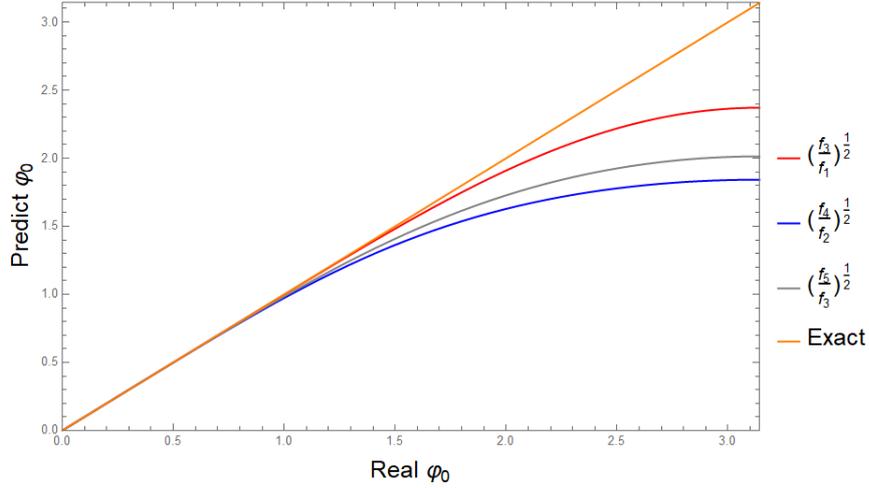

*Figure 2:* The predictions of $\varphi_0$ ($\varphi_i$) with different orders of derivatives.

The phenomenon of non-coincidence of the above-mentioned curves originates from the different accuracy of each derivative. Furthermore, to make the fractions in the scale sensor the same, we will reduce the accuracy of the lower-order derivatives on the denominator. Specifically, three sets of lower-precision derivative difference schemes are formed by points on different candidate stencils. In the end, the fractions formed by linear combination match the fraction of other different-order derivatives. The form of specific stencils and parameters are shown in the table 2.

*Table 2*

| $f_1$ | $f_{j-3}$ | $f_{j-2}$ | $f_{j-1}$ | $f_j$ | $f_{j+1}$ | $f_{j+2}$ | $f_{j+3}$ | $f_{j+4}$ |
|---|---|---|---|---|---|---|---|---|
| $a$ | $-\frac{3}{640}$ | $\frac{3}{128}$ | $-\frac{1}{192}$ | $-\frac{69}{64}$ | $\frac{141}{128}$ | $\frac{71}{1920}$ | | |
| $1-2a$ | | $-\frac{3}{640}$ | $\frac{25}{384}$ | $-\frac{75}{64}$ | $\frac{75}{64}$ | $-\frac{25}{385}$ | $\frac{3}{640}$ | |
| $a$ | | | $\frac{71}{1920}$ | $-\frac{141}{128}$ | $\frac{69}{64}$ | $\frac{1}{192}$ | $-\frac{3}{128}$ | $\frac{3}{640}$ |
| **$f_2$** | | | | | | | | |
| $b$ | $\frac{5}{48}$ | $-\frac{35}{48}$ | $\frac{19}{8}$ | $-\frac{67}{24}$ | $\frac{41}{48}$ | $\frac{3}{16}$ | | |
| $1-2b$ | | $-\frac{5}{48}$ | $\frac{13}{16}$ | $-\frac{17}{24}$ | $-\frac{17}{24}$ | $\frac{13}{16}$ | $-\frac{5}{48}$ | |
| $b$ | | | $\frac{3}{16}$ | $\frac{41}{48}$ | $-\frac{67}{24}$ | $\frac{19}{8}$ | $-\frac{35}{48}$ | $\frac{5}{48}$ |
| **$f_3$** | | | | | | | | |
| $c$ | $\frac{1}{8}$ | $-\frac{5}{8}$ | $\frac{1}{4}$ | $\frac{7}{4}$ | $-\frac{19}{8}$ | $\frac{7}{8}$ | | |
| $1-2c$ | | $\frac{1}{8}$ | $-\frac{13}{8}$ | $\frac{17}{4}$ | $-\frac{17}{4}$ | $\frac{13}{8}$ | $-\frac{1}{8}$ | |
| $c$ | | | $-\frac{7}{8}$ | $\frac{19}{8}$ | $-\frac{7}{4}$ | $-\frac{1}{4}$ | $\frac{5}{8}$ | $-\frac{1}{8}$ |

The thus constructed difference scheme $f_i$ is a function of the linear combination coefficients in its stencils. In particular, to prevent confusion, all the i-th order derivative difference schemes constructed with the highest accuracy of 8 candidate points in the following are denoted as $f_{ai}$. In order to make the fraction composed of each order derivative have the same scale prediction curve, the fitting formulas of a, b, and c are constructed through the mean inequality. The optimized coefficients are shown in Eq. (9).

$$sensor82: \quad a = 0.34286317405415073$$
$$b = -0.1407484382841254$$

$$sensor83: \quad a = 0.36137720814216684$$
$$b = -0.14261351265434885$$
$$c = 0.17566780549114422$$

(9)

Therefore, the curves of $\varphi_i$ (i = 1,2,3) - $\varphi_0$ will be very close to completely coinciding in value. However, the predicted-$\varphi_0$ does not coincide with the ideal $\varphi_0$. The method of piecewise fitting is used here. Specifically, this article takes sensor82 as an example, and the fitting function is derived in Eq. (10).

$$\text{sensor}82: \quad \varphi_{fix} = \max\left(\sum_{i=0}^{6} a_i (\varphi_{pre})^i, \ \pi\right)$$

| $\varphi_{pre}$ | $a_0$ | $a_1$ | $a_2$ | $a_3$ |
|---|---|---|---|---|
| [0, 0.5) | 0 | 0.9999998956451356 | $3.900270811075046e-6$ | -0.000051328665377752205 |
| [0.5, 1.0) | -0.000017888512305898928 | 1.0001786859466892 | -0.0007319531920454012 | 0.0015185947749004095 |
| [1.0, 1.5) | 0.011980579213337623 | 0.9366582236290302 | 0.13974773840006197 | -0.1647221146353264 |
| [1.5, 2.0) | 2.3941224915856445 | -7.776847147845627 | 13.441090514984964 | -11.014724950972207 |
| [2.0, 2.25) | 390.33987600308103 | -1137.9876058560228 | 1386.3052958543637 | -901.0507602974843 |
| [2.25, 2.4) | 38994.89410653587 | -102291.45001860967 | 111838.36745981683 | -65233.52575774523 |
| [2.4, +∞) | -450031.74260286614 | 740668.6745624893 | -381833.90461386944 | 1455.4616867523391 |

(10)

| $\varphi_{pre}$ | $a_4$ | $a_5$ | $a_6$ |
|---|---|---|---|
| [0, 0.5) | 0.0003248616784217192 | -0.0011047899889724334 | 0.0011861237502348357 |
| [0.5, 1.0) | -0.001437456093922423 | -0.0002183972196438273 | 0.0011005369199512049 |
| [1.0, 1.5) | 0.10967048933901359 | -0.040018235464370555 | 0.00707545265656572 |
| [1.5, 2.0) | 5.099430830157756 | -1.2671650441835973 | 0.13321770684773618 |
| [2.0, 2.25) | 329.92226176116696 | -64.5440392005354 | 5.273872456315239 |
| [2.25, 2.4) | 21410.20493680783 | -3749.1511770310144 | 273.6644583367362 |
| [2.4, +∞) | 63605.044277095556 | -20988.76107638564 | 2159.390591548757 |

The fitting function can be used to correct the reasonable former predicted results, which output the same value as the original scale. The fitting function can be considered as a correction to the predictable truncation error. The specific correction results are shown in figure 3,

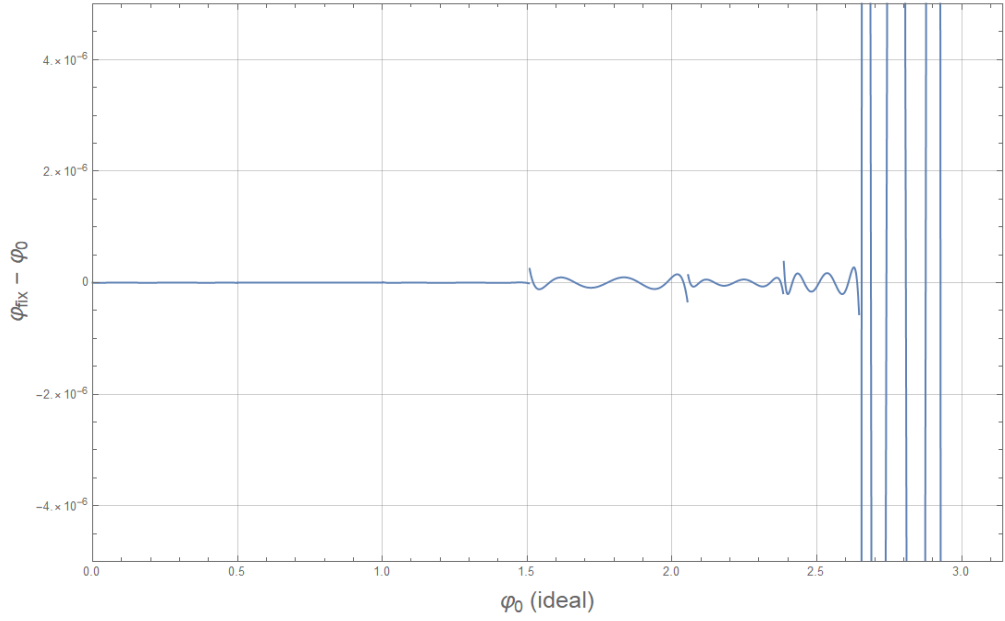

***Figure 3:*** The deviation between $\varphi_0$ and the fixed result.

In practical applications, in order to prevent small errors in the smooth region from causing the small-scale output of scale sensor, a threshold function needs to be added (Eq. 11) to the denominator of the scale sensor to eliminate the effect of the smooth region error.

$$sensor82core: \quad \varphi_{pre} = \sqrt{\frac{f_3 + f_4}{f_1 + f_2 + \varepsilon}} \tag{11}$$

$\varepsilon = 0.005 \cdot Exp[-500 \cdot max(f_{max} - f_{min},\ 0.01)]$ ($f$ is the candidate points in the stencils)

### 3.2. Smoothing of Scale Sensor

For the scale sensor sensor82, the output at different positions of a continuous-discrete function in the static test is not always continuous. The main reason is that there are absolute value functions in the derivative calculation function. This function is not continuous in the region where derivatives are nearly 0, that is, the result will oscillate near the phase of $\mathbb{Z}^*\pi/4$. To solve this problem, the derivative function part of the scale recognizer is modified (Eq. 12) so that the absolute value discontinuity does not appear in the derivative formula.

$$sensor82core(i): \quad \varphi_{pre} = \left( \frac{(f_3)^{2i} + (f_4)^{2i}}{(f_1)^{2i} + (f_2)^{2i} + \varepsilon} \right)^{\frac{1}{4i}} \tag{12}$$

The performance of the scale recognizer is compared statically, and the static test function (Eq. 13) is selected. The expression is as follows

$$TestFunction(x) = \sin[2x \cdot Exp(5x)] \tag{13}$$

Discrete the continuous functions mentioned above and select the grid width to be 1/193, The output of different scale sensors and the results of theoretical scales are shown in figure 4.

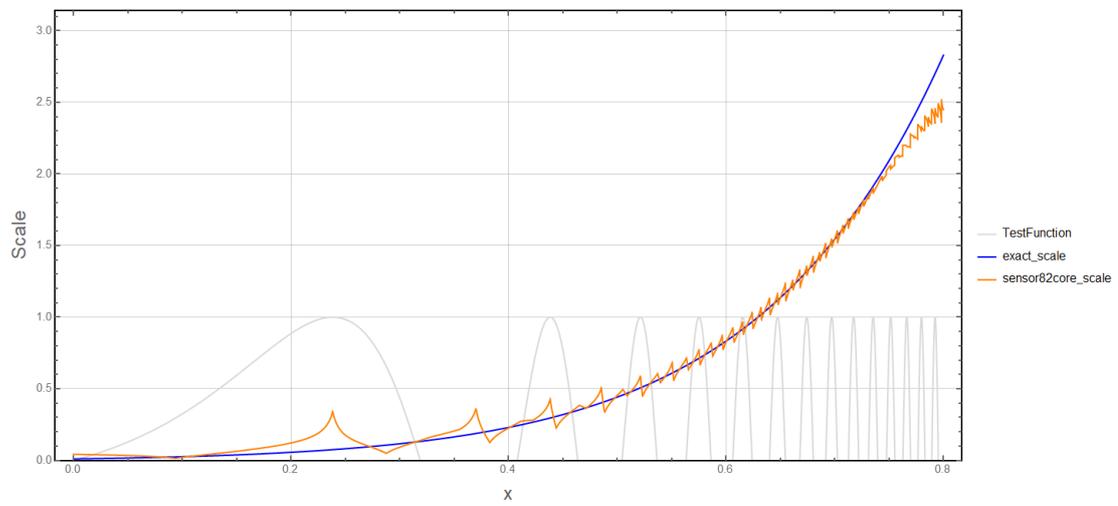

(a)

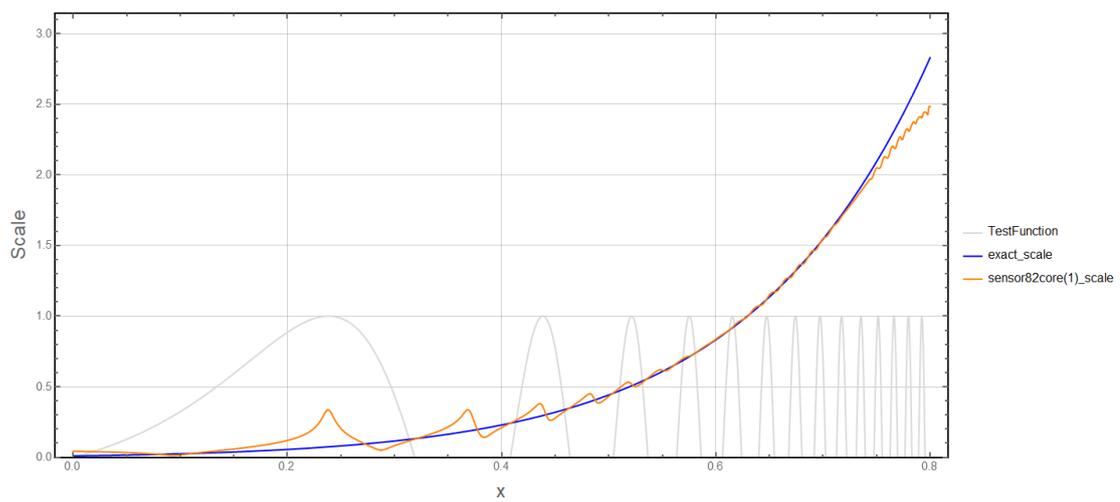

(b)

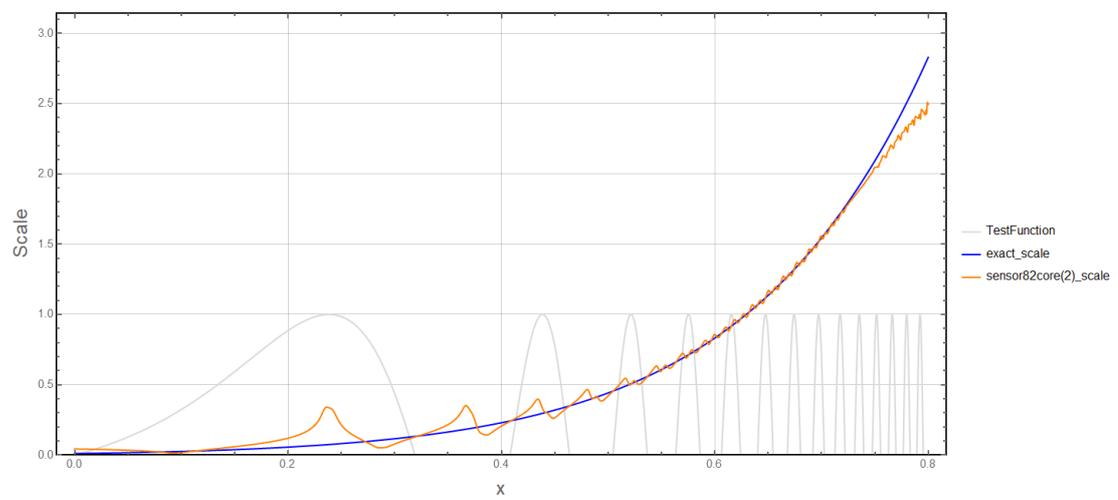

(c)

*Figure 4:* (a) Scales predicted by the sensor with absolute value function. (b) Scales predicted by the sensor with quartic function. (c) Scales predicted by the sensor with absolute square function.

The static test results show that the scale-position function derived by the square or higher-order power of the derivative term is more continuous than the absolute value derivative sensor, and there are no apparent peaks and discontinuities. This paper recommends using *sensor82core(1)*. Further, consider the correction function, similar to the advanced sensor82, the new correction function is optimized (Eq. 14),

$$\text{sensor}82[1]: \quad \varphi_{fix} = \max\left(\sum_{i=0}^{6} a_i (\varphi_{pre})^i, \pi\right)$$

| $\varphi_{pre}$ | $a_0$ | $a_1$ | $a_2$ | $a_3$ |
|---|---|---|---|---|
| $[0, 0.5)$ | 0 | 1.0000002011645452 | $-6.538772827759455e-6$ | 0.00007007386179275412 |
| $[0.5, 1.0)$ | $-0.010396868588218325$ | 1.0951249391596534 | $-0.3571269844978667$ | 0.7024167045788408 |
| $[1.0, 1.5)$ | 0.04345307270347291 | 0.7123016256681011 | 0.7590142365198556 | $-1.0324880461362562$ |
| $[1.5, 2.0)$ | 3.6316131287376257 | $-12.308591107617422$ | 20.33727924723695 | $-16.588590889092117$ |
| $[2.0, 2.25)$ | 488.1618745426508 | $-1415.8610178236356$ | 1714.6358490586008 | $-1107.6087576874504$ |
| $[2.25, 2.4)$ | 38674.63053016014 | $-101409.91256972586$ | 110829.33359080332 | $-64618.62787498283$ |
| $[2.4, +\infty)$ | $-56662.81099679229$ | 94779.42410072267 | $-49724.04800447986$ | 291.5965342770806 |

(14)

| $\varphi_{pre}$ | $a_4$ | $a_5$ | $a_6$ |
|---|---|---|---|
| $[0, 0.5)$ | $-0.0003199810688252983$ | 0.0005501295754577802 | $-0.00003429476510115172$ |
| $[0.5, 1.0)$ | $-0.7610006077526499$ | 0.428997014812631 | $-0.09764633135511866$ |
| $[1.0, 1.5)$ | 0.7678630067021806 | $-0.29766117001785825$ | 0.047885285162963385 |
| $[1.5, 2.0)$ | 7.620178068616666 | $-1.8713951817526557$ | 0.19314750324648547 |
| $[2.0, 2.25)$ | 402.8971077159796 | $-78.27177961384321$ | 6.348195560960004 |
| $[2.25, 2.4)$ | 21199.761322465405 | $-3710.793745374603$ | 270.7551183848493 |
| $[2.4, +\infty)$ | 8488.946383481727 | $-2852.1024282199796$ | 298.44459867651017 |

(In the following, only the Lid-driven cavity benchmark test uses *sensor82[1]*, while others still use *sensor82*)

## 4. Relationship between scale and dispersion and dissipation

This article constructs two formats, namely MDADdiss6th (hereafter also referred to as MDAD6th or MDADdiss) and MDADdisp6th (MDADdisp). The former only changes $\gamma_{diss}$, while $\gamma_{disp}$ is a constant, the latter is the opposite. The parameters $\gamma_{diss}(\varphi)$ and $\gamma_{disp}(\varphi)$ of these two schemes will be constructed separately below.

### 4.1. Dissipation function of MDADdiss6th

According to (Hu et al. 2012), the dispersion and dissipation relationship should satisfy $r(\varphi_0) = \left| \dfrac{\dfrac{\partial \Re(\varphi_0)}{\partial \varphi_0} - 1}{\Im(\varphi_0)} \right| \subset [1, 10]$, Through the above equation, the function of

dissipation-scale can be constructed (Eq. 15).

$$\gamma_{\text{diss}}(\varphi_0) = \frac{1}{r(\varphi_0)} \cdot \left| \frac{\frac{\partial \Re(\varphi_0)}{\partial \varphi_0} - 1}{g(\cos(\varphi_0))} \right| \qquad (15)$$

Where $g(\cos(\varphi_0)) = \dfrac{\Im(\varphi_0)}{\gamma_{\text{diss}}(\varphi_0)}$. If r={8, 9, 10}, $\gamma_{\text{disp}} = -0.01750475$, $\gamma_{\text{diss}}$ has the graph in figure 5.

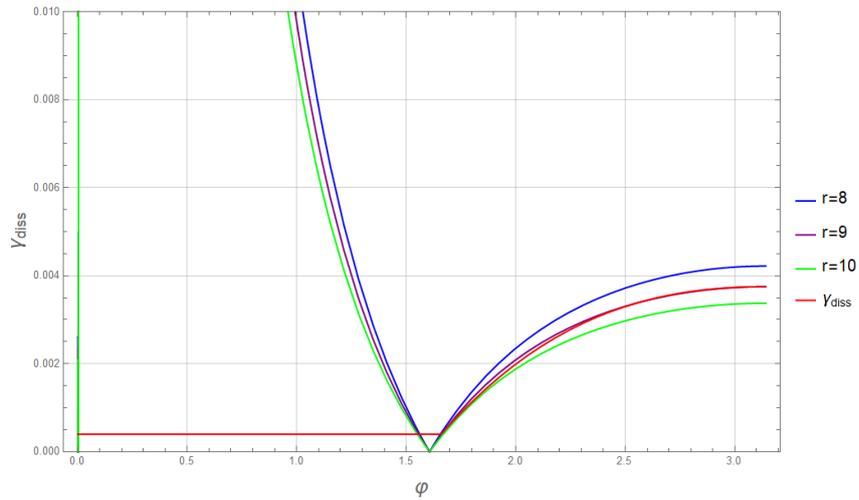

***Figure 5:*** The related dissipation function of scale with different r.

In order to obtain the expression of $\gamma_{diss}(\varphi)$ with a more stable r. A piecewise fitting is performed on $\gamma_{diss}(\varphi)$ (Eq. 16),

$$\gamma_{\text{diss}} = \begin{cases} 0.0004 & , \varphi < 1.66 \\ 0.00375 - \dfrac{0.00103 \cdot (\pi - \varphi)^3 + 0.001526 \cdot (\pi - \varphi)^2}{2} & , \varphi \geqslant 1.66 \end{cases} \qquad (16)$$

The curve is shown in the figure 5 above while $r(\varphi)$ calculated from this fitting function is shown in figure 6,

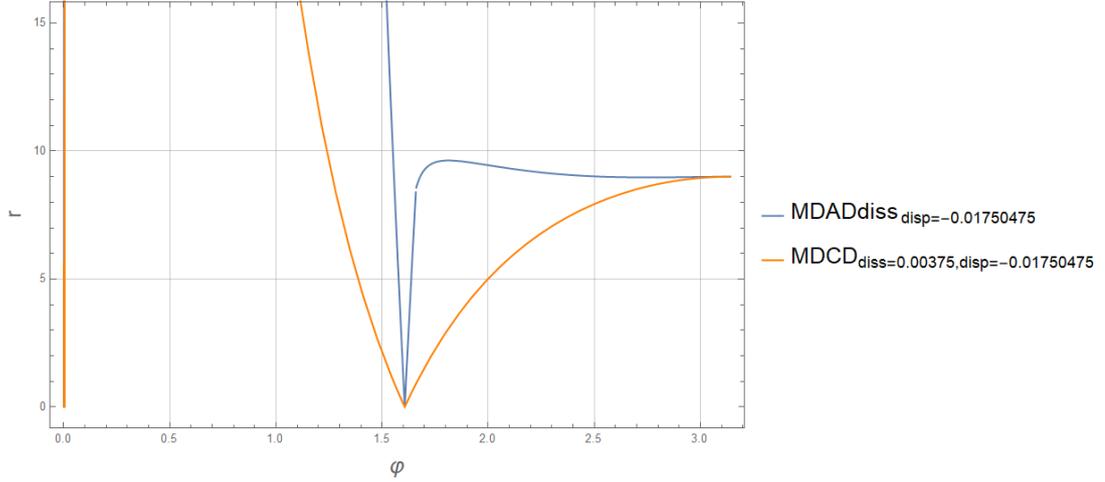

*Figure 6:* The equivalent r of MDADdiss and MDCD in different scales.

It can be seen that compared to the MDCD (Sun et al. 2014) the adaptive dissipation of smaller-scale regions does not have a slow rising process, but quickly approaches r=9.

### 4.2. Dispersion function of MDADdisp6th

The setting of the dispersion coefficient (Eq. 17) requires the smallest dispersion error, and practical only the solution of $\Re(\varphi_0) = \varphi_0$ are needed.

$$\gamma_{\mathrm{disp}} = \frac{\varphi_0 - \frac{3}{2}\sin(\varphi_0) + \frac{3}{10}\sin(2\varphi_0) - \frac{1}{30}\sin(3\varphi_0)}{-14\sin(\varphi_0) + 14\sin(2\varphi_0) - 6\sin(3\varphi_0) + \sin(4\varphi_0)} \quad (17)$$

This algorithm needs to be improved because of the poor robustness, and the following Euler and NS benchmark tests do not use this scheme.

## 5. Numerical results

### 5.1. Dispersion and Dissipation properties

Based on the linear convection equation, the ADR (Pirozzoli 2006) test is used to analyze the dispersion and dissipation characteristics of the scheme. The numerical solution will be represented by the Eq. (18).

$$\frac{\partial u}{\partial t} + a\frac{\partial u}{\partial x} = 0$$

$$\begin{cases} x \in (-\infty, +\infty) \\ u(0,x) = \mathrm{Exp}(ikx) \end{cases} \quad (18)$$

$$u(x,t) = \mathrm{Exp}\left(\mathrm{Im}(\Phi(\varphi))\,\frac{a \cdot t}{\Delta x}\right) \cdot \mathrm{Exp}\left(ik\left(x - \frac{\mathrm{Re}(\Phi(\varphi))}{\varphi} a \cdot t\right)\right)$$

The dispersion characteristic is represented by the $\mathrm{Re}(\phi(\varphi)) - \varphi$ curve (figure 7). It can be seen that for monochromatic waves, the MDADdisp scheme can achieve better spectral resolution by adaptively adjusting the dispersion coefficient. Its resolution limit will be reached near $\varphi = 2.9$. However, in terms of stability, due to the positive weights of the scheme's candidate stencils and the constrain of r, practically, the MDADdisp scheme needs to be improved in the application of other equations in order to achieve a more stable output.

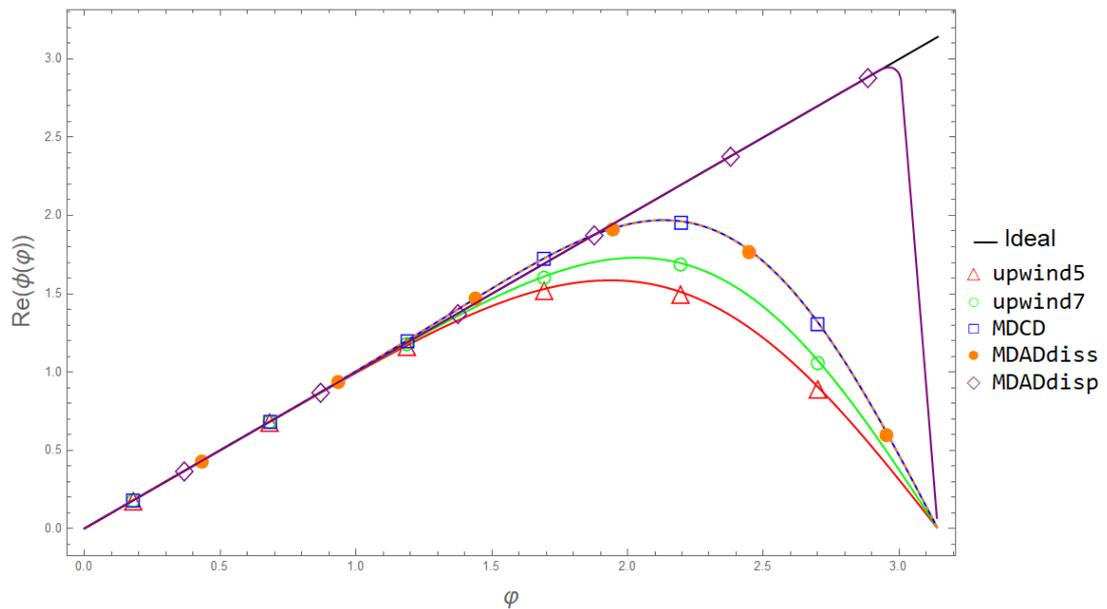

*Figure 7:* The dispersion properties of different schemes in ADR test

For the dissipation properties of the MDADdiss scheme (figure 8), due to the adaptive adjustment of the dispersion coefficient, compared with MDCD in the high spatial frequency area, the dissipative properties of MDAD can be rapidly enhanced. At the same time, the scheme can maintain very good low dissipation properties in the low spatial frequency region.

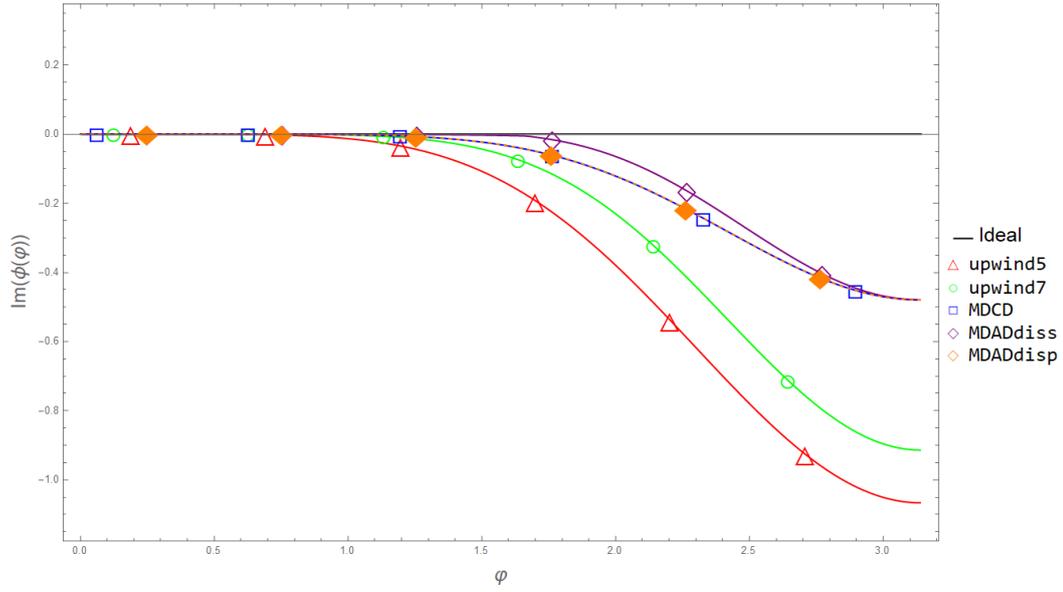

**Figure 8:** The dissipation properties of different schemes in ADR test

### 5.2. Convergence properties of the scheme

5.2.1. Linear convection equation

Consider the convergence rate of the solution of the linear convection equation, for periodic boundary conditions (Eq. 19).

$$\frac{\partial u}{\partial t} + a\frac{\partial u}{\partial x} = 0$$

$$\begin{cases} x \in [0,1] \\ u(0,x) = \frac{1}{m}\sum_{k=1}^{m}\sin(2\pi k x) \end{cases} \quad (19)$$

Applying different schemes, the relationships between the L2 error of the solution and the number of grids in the domain are presented in figure 9.

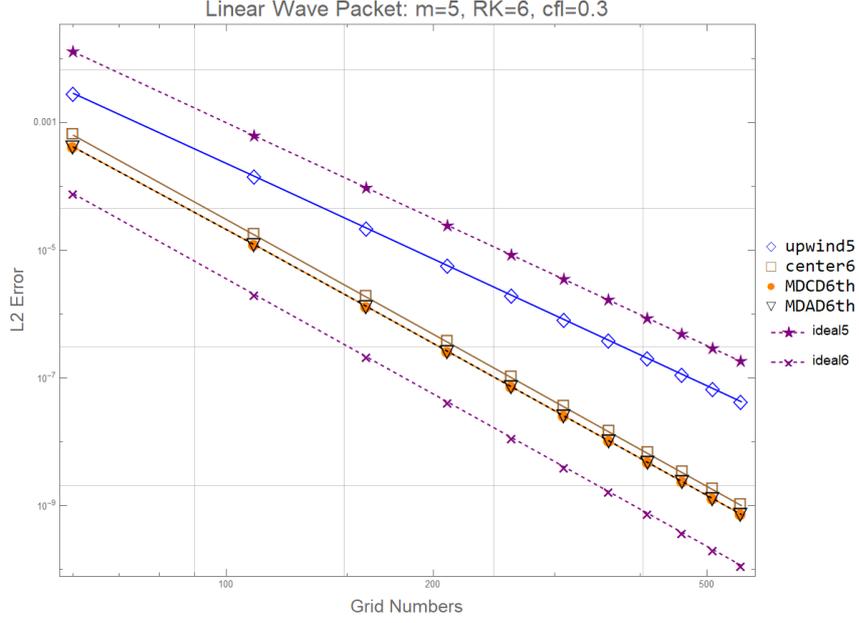

***Figure 9:*** The convergence properties of MCAD6th and other schemes in linear case.

The CFL coefficient is 0.3, and sixth-order Runge-Kutta scheme is utilized in the time integration (Luther 1968). As can be seen in the figure, in the continuous flow field, the solution of the linear partial differential equation of the MDAD6th scheme has the accuracy of a sixth-order scheme.

5.2.2. 2D-Vortex

In addition, consider the two-dimensional vortex case of an inviscid flow field. The initial conditions are as follows (Eq. 20),

$$\begin{cases} \rho(0,\vec{r}) = \left(1 - \frac{(\gamma-1)\cdot\varepsilon^2}{8\pi^2\gamma}\cdot\mathrm{Exp}\left(1-r^2\right)\right)^{\frac{1}{\gamma-1}} \\ u(0,\vec{r}) = 1 - \frac{\varepsilon\vec{r}\cdot\mathbf{e_y}}{2\pi}\cdot\mathrm{Exp}\left(\frac{1-r^2}{2}\right) \\ v(0,\vec{r}) = 1 + \frac{\varepsilon\vec{r}\cdot\mathbf{e_x}}{2\pi}\cdot\mathrm{Exp}\left(\frac{1-r^2}{2}\right) \\ p(0,\vec{r}) = \rho(0,\vec{r})^\gamma \end{cases} \quad (20)$$

Take the region of space as $[0,20]\otimes[0,20]$ with periodic boundary conditions in both directions, where the $\vec{r}$ of the initialization condition is the position vector from the geometric center of the domain to the target point. The CFL coefficient is 0.6, and sixth-order Runge-Kutta scheme is utilized in the time integration. Considering that the period of the flow field is 20s, take the L2 error of the initial state and the solution at 20s as the criterion of convergence.

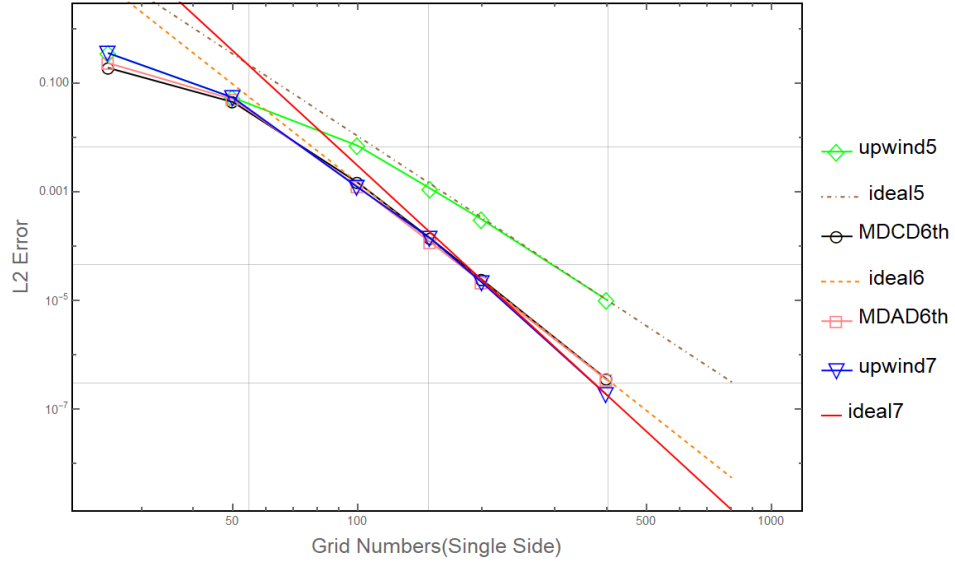

**Figure 10:** The convergence properties of MCAD6th and other schemes in non-linear case.

As can be seen in figure 10, in the continuous flow field, the solution of the non-linear equation of the MDAD6th scheme also has the accuracy of a sixth-order scheme.

**5.3. 2D Taylor-Green Vortex**

The initial conditions are set as follows (Eq. 21)

$$\begin{cases} \rho(0,x,y) = 1 \\ u(0,x,y) = \sin\left(2\pi\dfrac{x}{L}\right)\cos\left(2\pi\dfrac{x}{L}\right) \\ v(0,x,y) = -\cos\left(2\pi\dfrac{x}{L}\right)\sin\left(2\pi\dfrac{x}{L}\right) \\ p(0,x,y) = \dfrac{\rho(0,x,y)}{24}\cdot(\cos\left(4\pi\dfrac{x}{L}\right) + \cos\left(4\pi\dfrac{y}{L}\right)) + 100; \end{cases} \quad (21)$$

Take the region of space as $[0,L]\otimes[0,L]$ ($L=1.0$) with periodic boundary conditions in both directions. The results are shown is figure 11,

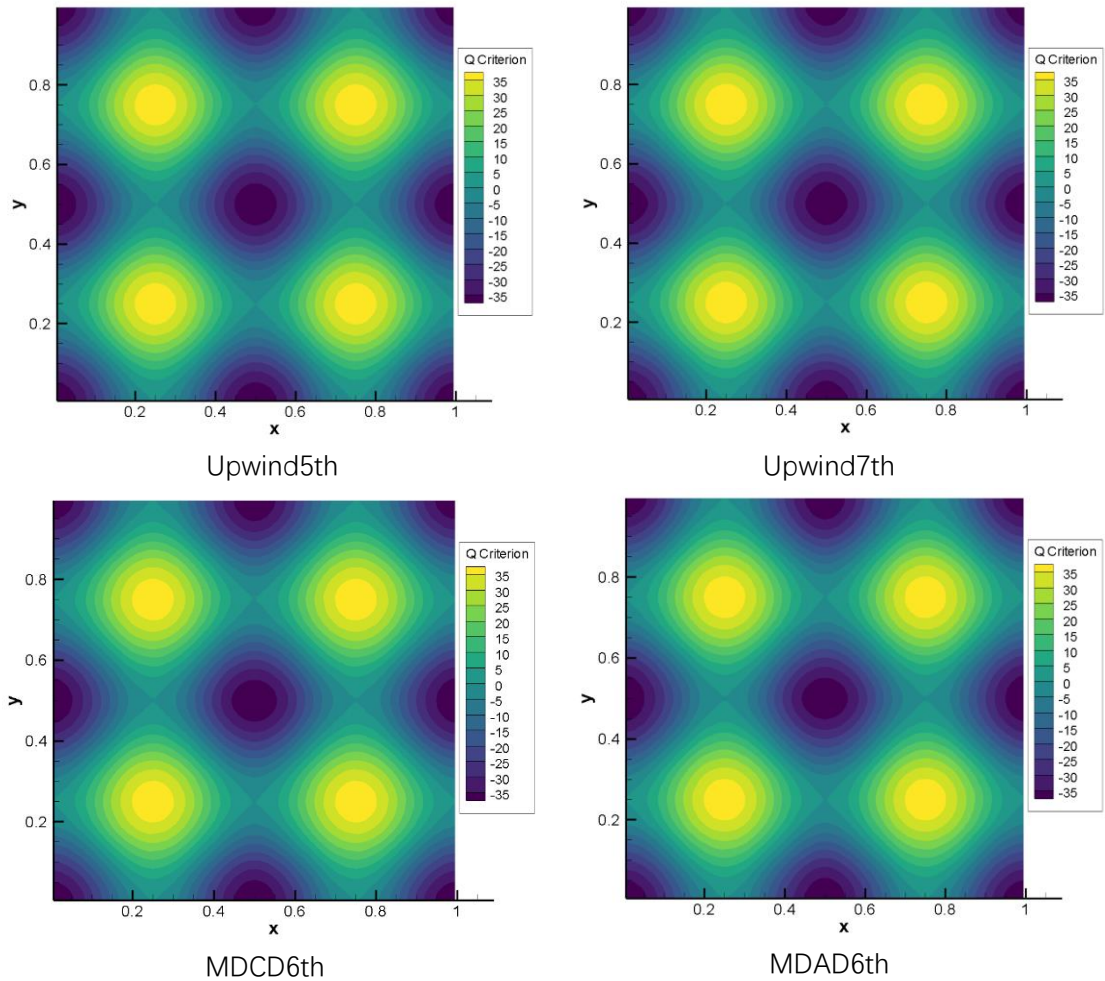

Upwind5th
Upwind7th
MDCD6th
MDAD6th

*Figure11:* Q criterion (Mesh 100*100, CFL=0.3, Time=25.0s)

There are no obvious differences between the solution of the Euler equations. Considering the NS equation with a mesh of $150 \otimes 150$, CFL = 0.6, applying fourth-order Runge-Kutta scheme and seventh-order upwind scheme. Figure 12 represents the result at time= 300s and Re=3000.

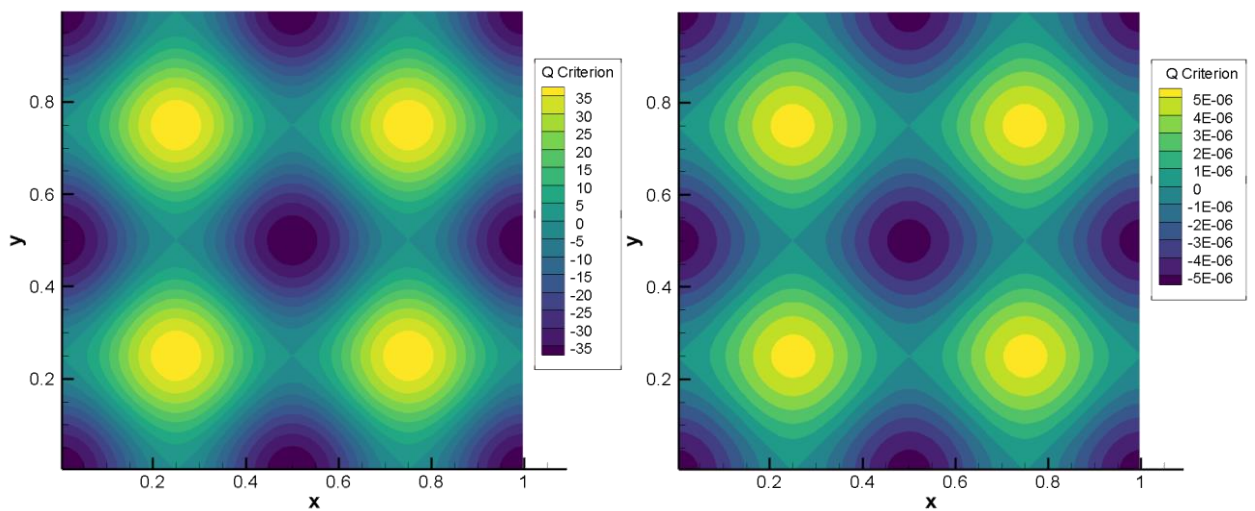

*Figure 12:* Q criterion (grid number 150*150, Time=Left:0.0s Right:300.0s)

As a result, there is still no small-scale structure. According to (Sengupta, Sharma, and Sengupta 2018), the vortex in a region with $200 \otimes 200$ mesh and local refinement can derive small-scale results as shown in the figure 13. Therefore, the reason that there is no small-scale structure in the results is that the grids are too sparse. Since there is not enough computing power for more accurate results and the time consumption is too large, resolved results of this case are yet to be calculated.

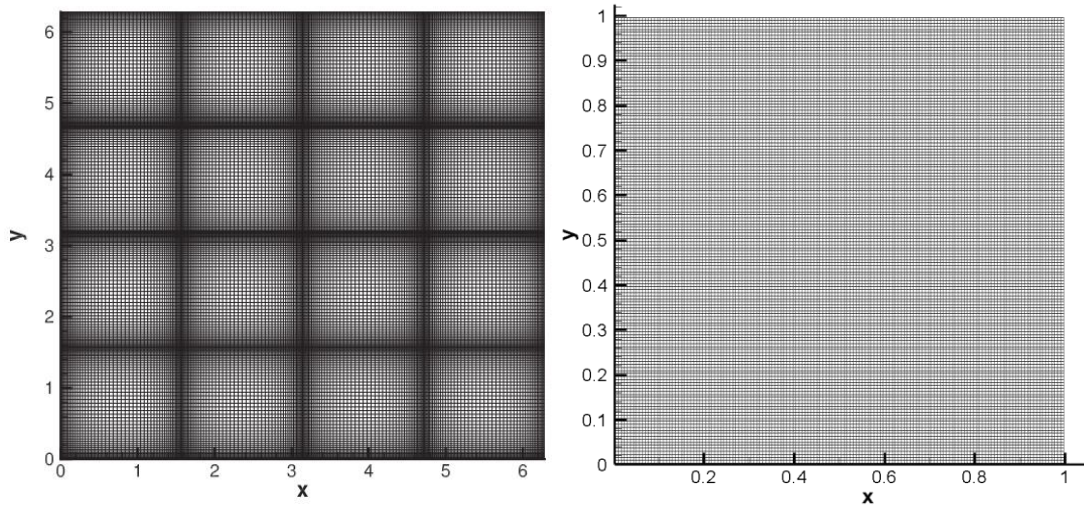

*Figure 13:* Left: Sengupta's Mesh. Right: Mesh in this Article

### 5.4. 2D Lid-driven cavity problem

Set density of the fluid $\rho = 1.0$ and $U_0 = 1.0$. For the setting of boundary conditions, the virtual grids outside the domain are mirrored corresponding to the values inside the boundary. Readers can refer to figure 14 for the specific boundary orthogonal.

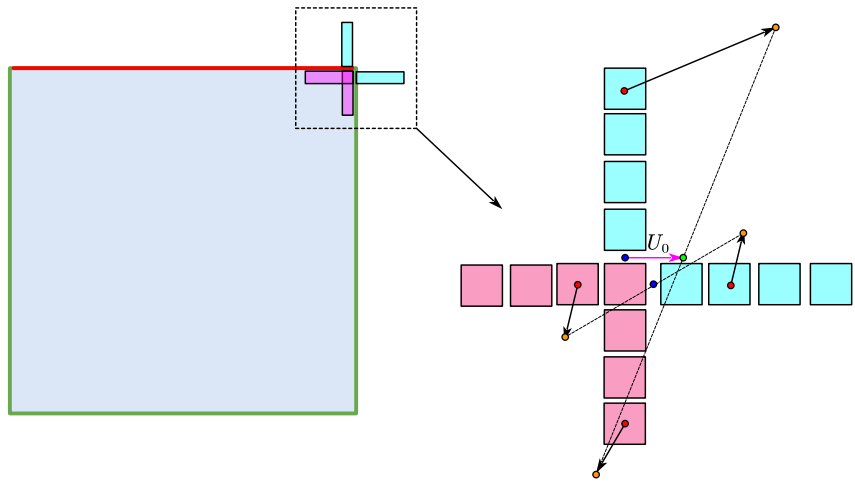

*Figure 14:* The setting of Boundary Condition in 2D Lid-driven cavity problem

In this case, the mesh of the domain is $50 \otimes 50$ with Re=1000 and a fourth-order Runge-Kutta method is applied until the flow field reaches a steady solution. Although this number of grids is not sufficient to calculate the field of Re=1000, this case aims to illustrate the problem of the MDAD scheme at the boundary.

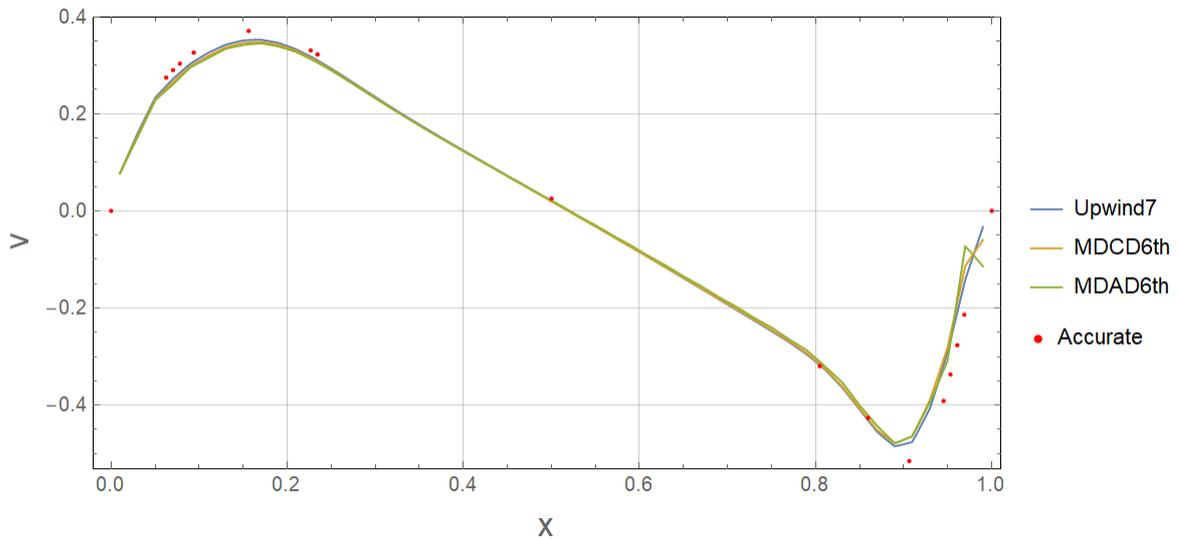

*Figure 15:* The x-v graph at the center of the domain parallel to the boundary (y=0.5). The exact solution refers to (Ghia, Ghia, and Shin 1982).

Noted that non-physical solution (figure 15) appears on the right boundary of MDAD6th. For MDCD6th, although the right boundary still has non-physical components compared with Upwind7, it is significantly better than MDAD6th. If the value of the dissipation function's output in the MDAD6th scale sensor is artificially increased, a relatively physical solution can still be obtained. Based on this, we speculate that this non-physical result is due to a greater dissipation (instead of the dissipation in MDAD6th) required by the right boundary condition. The problem is no longer obvious in larger grid numbers' cases, but it still exists. Meanwhile, the low dissipation of the smooth area does not significantly optimize the solution of the corresponding region which is still caused by the inaccurate calculation of the boundary and so that the advantage of MDAD is concealed.

For a mesh of $100 \otimes 100$, Re=1000 and with a fourth-order Runge-Kutta scheme, we calculate till a steady-state solution is reached. The x-v relationship in each scheme can obtain almost coincident accurate results. However, there are still non-physical oscillations in the right boundary of the MDAD flow field. In order to alleviate the influence of this problem, 4 grids of dissipation coefficient corresponding to small-scale results are added to the right area, that is, MDCD6th is applied in this stripe near the right boundary instead. This algorithm is tentatively named MDsemiAD6th. Figure16 show the difference of dissipation in the region.

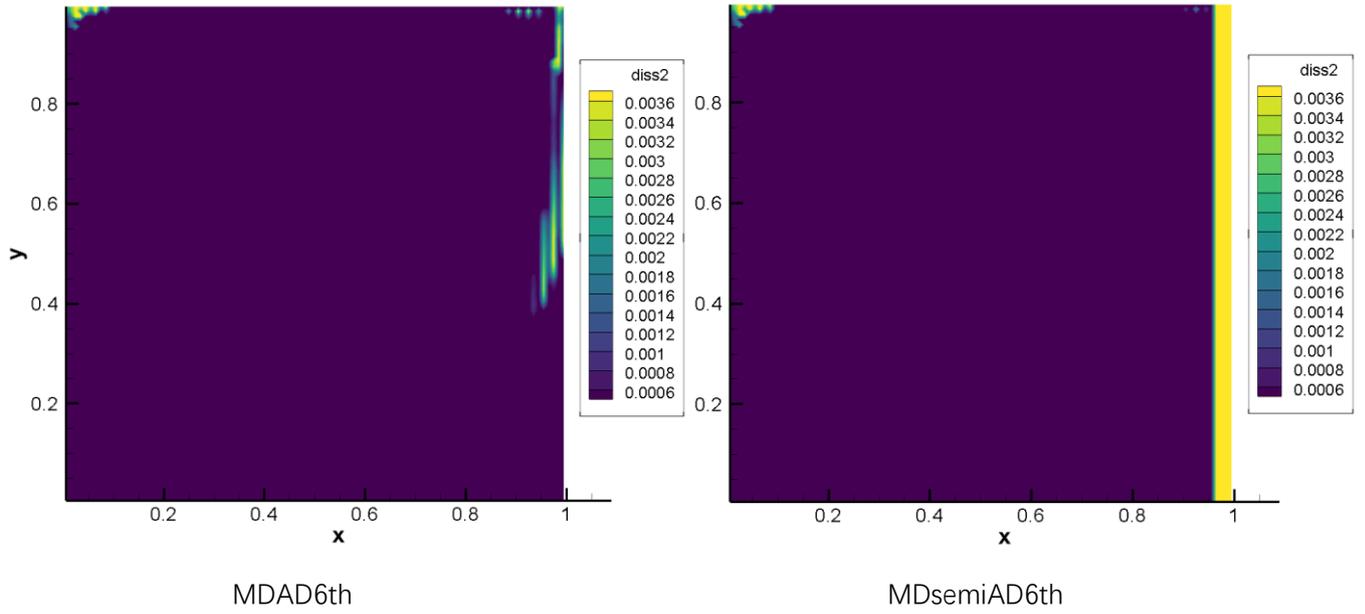

MDAD6th          MDsemiAD6th

*Figure 16:* The dissipative value corresponding to the scale of G vector (flux in y-direction) after diagonalization and matrix splitting transformation.

Further, the results of v in the domain for using different schemes are shown in the figure 17.

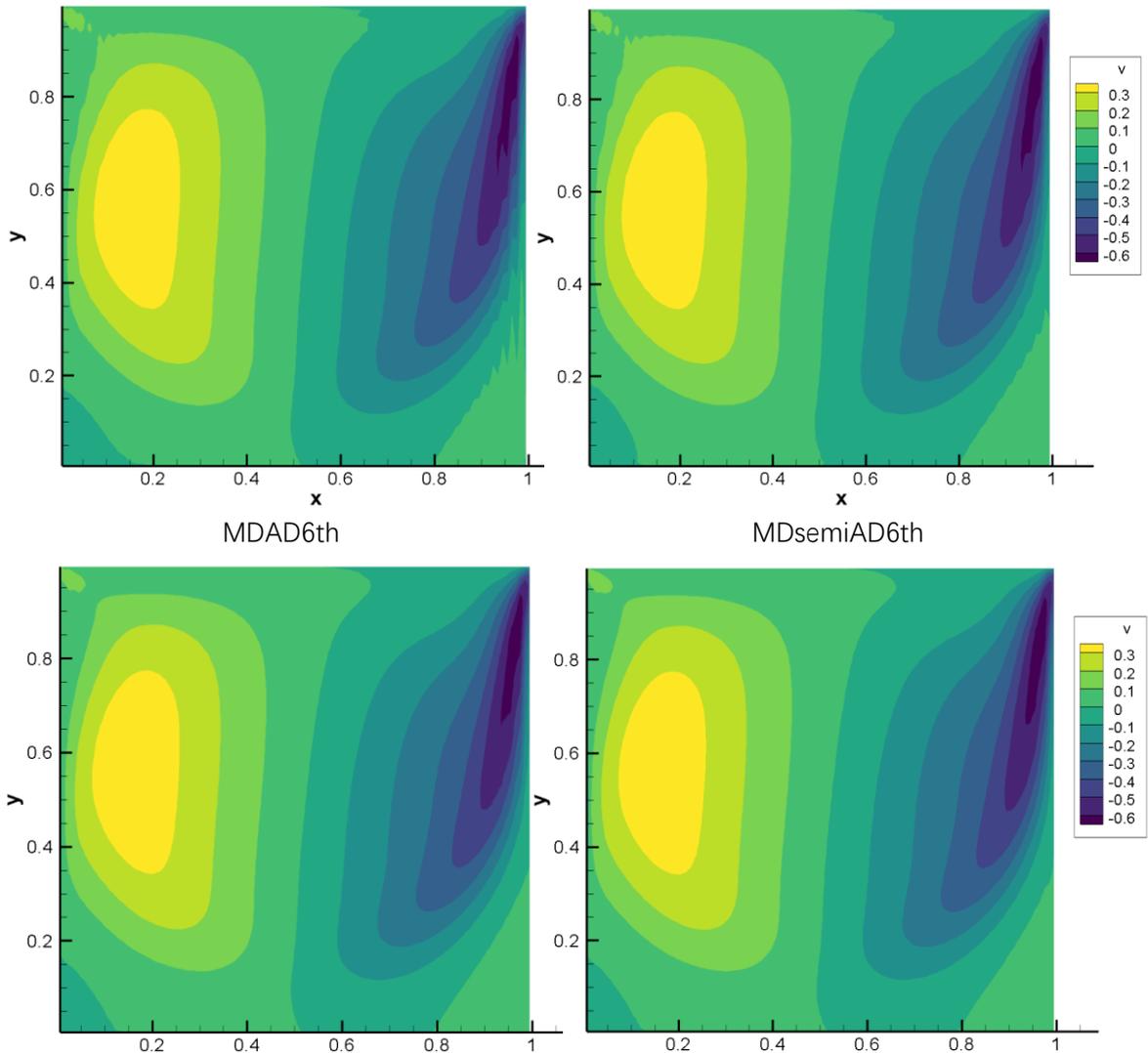

MDAD6th          MDsemiAD6th

MDCD6th                                         UpWind7th

*Figure 17:* The y-velocity contour plot in the region of different schemes

Through the analysis of the above results, it can be seen that only relying on the dissipation calculated by the MDAD scale sensor is insufficient for the suppression of non-physical oscillations in the area near the right boundary. To some extent, it shows that Hu's dispersion-dissipation relationship is not completely applicable to the wave packet propagation and dissipation assumptions for the wall boundary conditions. The fields of MDCD6th and UpWind7th are relatively smooth. The MDsemiAD6th, which has been artificially added with high dissipation, performed better on the right boundary, but at the same time, it still performed slightly worse than MDCD6th. This benchmark case puts forward a new algorithm for the use of MDAD6th and its scale sensor.

## 6. A new perspective of the utilization of scale and adaptive ability

Considering the variable dispersion format (MDAD6thdisp) of the MDAD6th, the semi-discrete scheme corresponding to different dispersion coefficients on different sides of the interface may have positive dissipation properties which makes the results prone to instability. In addition, for the selection of the dissipation coefficient, the dissipation coefficient derived by the scale sensor may not completely restrain the oscillation caused by the high-order derivative discontinuity due to the imperfect boundary conditions.

For the former, we can use two different scales with two sets of dissipation coefficients on both interfaces of the grid to calculate the flux stably. In this way, the difference of flux between the two sides of the interface can be calculated with the same scale, to avoid the problem of subtraction of flux between different scales. However, this method has the problem of low computational efficiency. Almost twice the amount of computation is needed in this problem. At the same time, for the high computational complexity of the MDAD6th scale sensor, compared with MDCD6th, it can be found that the time consumption of the same case of MDAD6th is about 1.5-2 times higher than that of MDCD6th.

The new idea is inspired by the appliance of MDsemiAD6th in the Lid-driven cavity problem. The core idea is to use the corresponding optimal MDCD6th scheme in a divided domain. In a sense, MDAD6th is an extreme occasion of this algorithm that the domain is divided into units with only one single grid. For example, for the following situations in the Lid-driven cavity problem (figure 18), the domain can be divided into three different types of regions, corresponding to different MDCD coefficients.

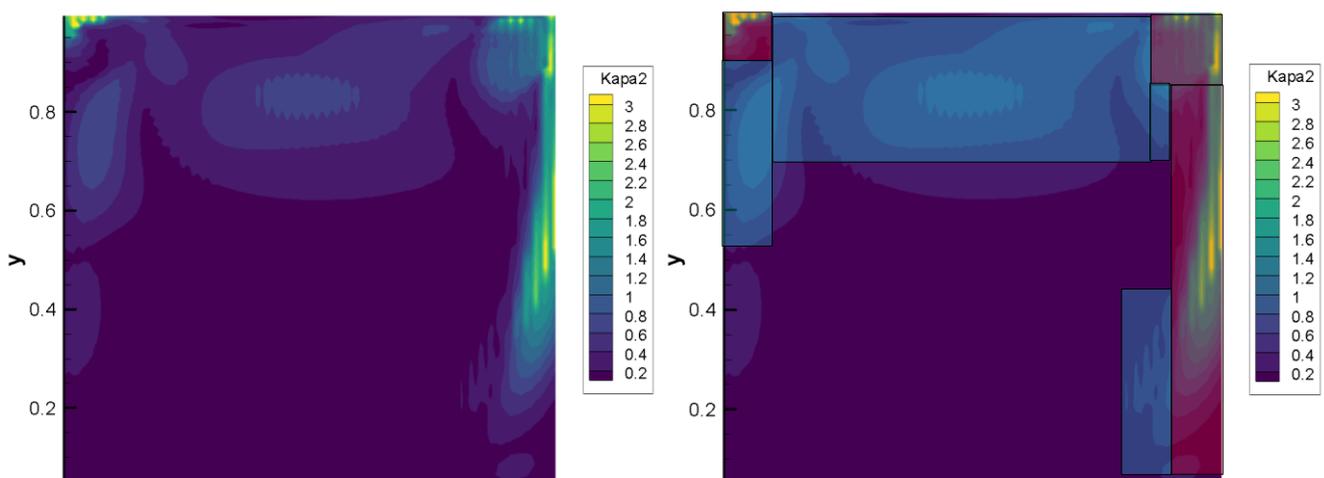

**Figure 18:** Left: Scale Sensor output in Lid-driven cavity problem. Right: Schematic diagram of the division of the domain

The area division process can be an adaptive procedure that is performed in every time step. At the same time, the area division algorithm doesn't need to apply the scale sensor to calculate the scale for each grid by judging the area's scale by sampling. For multi-scale problems in practical applications, the divided area can be larger which is because the spectral properties of the domain where the scale is identified and optimized for each grid are unstable and meanwhile a larger divided area represents fewer area connections means fewer repeated flux calculations. The area division algorithm is yet to be developed, and it seems that it has usefulness and advantages for practical problems such as jet flow.